\documentclass[a4paper]{article}
\usepackage{multirow}
\usepackage{cite}
\usepackage{INTERSPEECH2020}

\title{Deep MOS Predictor for Synthetic Speech Using Cluster-Based Modeling}
\name{Yeunju Choi, Youngmoon Jung, Hoirin Kim}
\address{
  School of Electrical Engineering, KAIST, Daejeon, Republic of Korea}
\email{\{wkadldppdy,dudans,hoirkim\}@kaist.ac.kr}

\begin{document}

\maketitle

\begin{abstract}
While deep learning has made impressive progress in speech synthesis and voice conversion, the assessment of the synthesized speech is still carried out by human participants. Several recent papers have proposed deep-learning-based assessment models and shown the potential to automate the speech quality assessment. To improve the previously proposed assessment model, MOSNet, we propose three models using cluster-based modeling methods: using a global quality token (GQT) layer, using an Encoding Layer, and using both of them. We perform experiments using the evaluation results of the Voice Conversion Challenge 2018 to predict the mean opinion score of synthesized speech and similarity score between synthesized speech and reference speech. The results show that the GQT layer helps to predict human assessment better by automatically learning the useful quality tokens for the task and that the Encoding Layer helps to utilize frame-level scores more precisely.
\end{abstract}

\section{Introduction}

Recent advances in deep learning have led to significant growth in various speech processing fields \cite{Han2019, WaveNet, Tacotron2, TransformerTTS}.
However, in contrast to speech recognition task, there is no ``right answer'' in speech generation tasks such as text-to-speech (TTS) or voice conversion (VC).
For this reason, subjective measures such as the mean opinion score (MOS) and similarity score have been used to evaluate naturalness and similarity, respectively \cite{King}.
That is, the quality measurement of the synthesized speech is still carried out by many human subjects, which is expensive and time-consuming \cite{Eval}. Moreover, the results may change depending on several factors, such as human subjects and audio hardware.

There are many objective measures of speech quality to reflect human perception \cite{MCD, PESQ, ANIQUE, P563}. 
The most widely used measures are Mel-cepstral distance (MCD) \cite{MCD} and the perceptual evaluation of speech quality (PESQ) \cite{PESQ}. 
However, these are full-reference measures in that they need ground-truth speech as reference. 
There are also no-reference measures such as ANIQUE \cite{ANIQUE} or ITU-T recommendation P. 563 \cite{P563}. 
However, most of these measures are targeted at detecting artifacts caused by lossy compression and transmission in telephony, not at evaluating the quality of synthetic speech.

With the advances in deep learning, researchers have recently proposed deep-learning-based models that can evaluate the quality of synthetic speech without reference speech \cite{HierarchicalAssessment, AutoMOS, QualityNet, MOSNet}. Patton \textit{et al.} \cite{AutoMOS} proposed AutoMOS, based on long short-term memory (LSTM), to predict MOS values. Fu \textit{et al.} \cite{QualityNet} proposed Quality-Net based on bidirectional LSTM (BLSTM) to predict the frame-level PESQs. Recently, Lo \textit{et al.} \cite{MOSNet} proposed MOSNet that generates frame-level MOSs from the features of convolutional neural network-BLSTM (CNN-BLSTM) and predicts the utterance-level MOS using the frame-level scores.  
Moreover, they modified MOSNet to evaluate the similarity score for VC and extended the deep-learning-based quality assessment area to similarity score prediction.

These researches have shown the potential to automate the assessment of the synthesized speech using deep neural networks. 
However, it is difficult to understand the assessment criteria for speech quality evaluation performed by humans. 
Everyone has a different perspective on speech quality, and even the same person has a different perspective each time. Therefore, we propose a model using a global quality token (GQT) layer that can automatically learn the criteria as soft clusters.

Furthermore, even though Quality-Net \cite{QualityNet} and MOSNet \cite{MOSNet} showed performance improvement by assuming that the quality score of an utterance is an average of frame-level scores, the exact relation between the utterance- and frame-level scores remains poorly understood. 
With the inspiration that humans will determine an utterance-level score in a more sophisticated way, we propose a model using an Encoding Layer to aggregate frame-level scores by considering not only the simple average of the frame-level scores, but also their distribution information.

\section{Relation to prior work}
\label{sec:relation_to_prior_work}

Lo \textit{et al.} \cite{MOSNet} proposed and compared three different model architectures for MOS prediction. All of them consist of a feature extractor, two fully-connected (FC) layers, and a global average pooling (GAP) layer. The main difference between the three models is the architecture of the feature extractor; CNN, BLSTM, and CNN-BLSTM. 
In this paper, we use the CNN-BLSTM-based MOSNet as our baseline, which showed the best performance among them. The architecture of MOSNet is shown in Table \ref{tab:architecture}. 
First, the feature extractor generates frame-level feature vectors from an input magnitude spectrogram. 
Then, the following FC layers map each frame-level feature vector into a frame-level score.
Finally, the GAP layer outputs the utterance-level score by averaging the frame-level scores.
They formulate a loss function using both utterance-level mean squared error (MSE) and frame-level MSE as follows:
\begin{equation}
L = \frac{1}{S}\sum_{s=1}^S[(\hat{Q}_s-Q_s)^2+\frac{\alpha}{T_s}\sum_{t=1}^{T_s}(\hat{Q}_s-q_{s, t})^2]\,,
\end{equation}
where S is the number of training utterances, and $T_s$ is the number of the frames of the \textit{s}-th utterance. $\hat{Q}_s$ and $Q_s$ are the ground-truth and predicted value for the utterance-level score of the \textit{s}-th utterance, respectively.
$q_{s, t}$ is the predicted frame-level score at time \textit{t} for the \textit{s}-th utterance, and $\alpha$ is a weighting factor for the frame-level MSE.

Furthermore, they extended CNN-based MOSNet to predict the similarity score between a pair of utterances. Two utterances of an input pair have the same length by zero-padding and share the convolution layers among the modified CNN-based MOSNet. 
Two CNN feature maps from the input pair are concatenated and fed into the following FC layers and a GAP layer to generate a similarity score.
In this work, we propose SIMNet by modifying CNN-BLSTM-based MOSNet in a similar way, and used this as a baseline model for similarity score prediction.
SIMNet concatenates two different feature maps of the shared CNN from the utterance pair and uses the result as an input of the following BLSTM layer. 

Two main contributions of our work are the following.
First, motivated by a global style token (GST) \cite{GST}, we propose a model using a global quality token (GQT) layer which learns the tokens that reflect the criteria for speech quality evaluation.
The GQTs operate in the same way as the GSTs, but we call them global quality tokens because they are learned in terms of speech quality. 
Second, we propose a model using an Encoding Layer which aggregates the frame-level scores by considering the distribution of them. Although Quality-Net and MOSNet showed that using frame-level scores improves MOS prediction performance, they only considered average information of the frame-level scores. We consider not only average information but also distribution information of frame-level scores.

\section{Proposed models}
\label{sec:proposed_models}
To improve MOSNet, we propose two models based on the GQT layer and the Encoding Layer, respectively. 
Throughout this paper, `+ GQT' and `+ EL' stand for using the GQT layer and using the Encoding Layer, respectively. Our third model is a combination of the first and second models.
For MOSNet, we use 16, 16, 32, and 32 filters for each convolutional block containing three convolutional layers.
The architectures of the four models are described in Table \ref{tab:architecture}.

\begin{table}[t]
\caption{Configuration of the model architectures. + GQT and + EL denote using the GQT layer and the Encoding Layer, respectively. The convolutional layer parameters are denoted as conv\{receptive field size\}-\{number of channels\}/\{stride\}. N is the number of frames. SC is skip connection. GAP is global average pooling. K is the number of the codewords for the Encoding Layer. For + EL and + GQT + EL, an additional FC layer of a pooling layer is omitted for simplicity.}
\label{tab:architecture}
\vspace{-0.5cm}
\begin{footnotesize}
\begin{center}
\renewcommand{\tabcolsep}{1.1mm}
\renewcommand{\arraystretch}{1.1}
\begin{tabular}{c|c|c|c|c}
\hline
model & \textbf{MOSNet} & \textbf{+ GQT} & \textbf{+ EL} & \textbf{+ GQT + EL} \\ 
\hline\hline
input & \multicolumn{4}{c}{N$\times$257 magnitude spectrogram} \\ \hline
& \multicolumn{4}{c}{} \\
\multirow{2.0}{*}
{\begin{tabular}[c]{@{}c@{}}conv.\\ layers\end{tabular}}
& \multicolumn{4}{c}{$\left\{ \begin{array}{cc} conv3 - (channels)/1
                        \\ conv3 - (channels)/1
                        \\ conv3 - (channels)/3
                        \end{array}\right\}$ $\times$ $4$}  \\  
& \multicolumn{4}{c}{} \\
& \multicolumn{4}{c}{$channels = [16, 16, 32, 32]$} \\
\hline
\begin{tabular}[c]{@{}c@{}}GQT layer\\with SC\end{tabular} 
& - &  \# GQTs = 10 & - &  \# GQTs = 10 \\
\hline
\begin{tabular}[c]{@{}c@{}}recurrent\\ layer\end{tabular} & \multicolumn{4}{c}{BLSTM-128} \\
\hline
\multirow{2}{*}{\begin{tabular}[c]{@{}c@{}}FC\\ layers\end{tabular}}
& \multicolumn{4}{c}
{\begin{tabular}[c]{@{}c@{}}FC-128,\\ ReLU,\\ dropout\end{tabular}} \\ \cline{2-5} 
& \multicolumn{4}{c}{FC-1 (\textbf{\textit{frame-level scores}})} 
\\ \hline
\begin{tabular}[c]{@{}c@{}}pooling\\ layer\end{tabular}
& \multicolumn{2}{c|}{\begin{tabular}[c]{@{}c@{}}GAP layer \\ (\textbf{\textit{utterance-level score}})\end{tabular}} 
& \multicolumn{2}{c}
{\begin{tabular}[c]{@{}c@{}}Encoding Layer (K=10) \\ \& GAP layer \\ (\textbf{\textit{utterance-level score}})\end{tabular}} \\ 
\hline
\end{tabular}
\end{center}
\vspace{-0.6cm}
\end{footnotesize}
\end{table}

\subsection{Global quality tokens for MOS and similarity score}
A GST model is proposed in \cite{GST} for style-expressive end-to-end TTS. It consists of a reference encoder, style token layer, and sequence-to-sequence model. 
The reference encoder extracts reference embedding of the reference utterance to refer the style.
The GSTs of the style token layer, which are shared by all training sequences, become the soft clusters of reference embedding through training. 
The goal of the style token layer is to calculate the style embedding, which represents the style of the reference utterance, as a weighted sum of the GSTs. The weights assigned to the GSTs are learned by a multi-head attention module \cite{AttentionIsAllYouNeed}. 
The style token layer consisting of GSTs and the attention module is randomly initialized and then trained jointly with the whole GST model. 
Therefore, the GSTs can become useful soft clusters for style modeling, and the GST model can synthesize speech with a specific style of the reference utterance by extracting the style embedding from it.

Although Wang \textit{et al.} \cite{GST} suggested the GST layer for TTS, they also showed that GSTs could be used as speaker classification features. 
There is also a report \cite{ImprovingUnsupervisedStyleTransfer} that the GST layer helps to improve speech recognition performance. 
In this paper, we report for the first time that the GST layer also helps to improve the performance of the quality assessment task. Here, we refer to the GST as the GQT, as we mentioned in the previous section. Besides, the reference encoder, GQTs, and multi-head attention module are collectively called the GQT layer. 

As the GSTs are targeted for a separate reference utterance, they need a separate reference encoder. 
Unlike GSTs, the GQTs are targeted for the same input utterance of MOS prediction.
Therefore, we design our reference encoder to share the convolutional layers with MOSNet. A gated recurrent unit (GRU) \cite{GRU} layer follows the convolutional layers and the last hidden state of the GRU layer serves as the reference embedding.
Then, the quality embedding is calculated in the same way to calculate the style embedding in \cite{GST}. 
As a kind of skip connection, the quality embedding is added to all the frame-level feature vectors of the CNN.
We use the resulting representation as an input of the BLSTM layer. 

When we use a GQT layer for a similarity score prediction task, we apply the shared GQT layer to generate two quality embeddings from an utterance pair. Then we add each quality embedding to the corresponding CNN feature map before the concatenation of two CNN feature maps.

\subsection{An Encoding Layer for MOS and similarity score}
The Encoding Layer \cite{DeepTEN} was proposed for texture recognition by learning the inherent visual codewords directly from the loss function.
The codewords are learned from the distribution of the CNN features. The Encoding Layer also acts as a pooling layer, which converts feature vectors of any size into a fixed-length representation. Given $N$ feature vectors, $X = \{x_1,...,x_N\}$, and $K$ codewords, $C = \{c_1,...,c_K\}$, the output representation of the Encoding Layer is $e = \{e_1,..., e_K\}$, called residual encoding vector. The residual vector $r_{ik}$ is calculated by $r_{ik} = x_i - c_k$ and the assigning weight for $r_{ik}$ is given by
\begin{equation}
w_{ik} = \frac{\exp(-s_{k}\|r_{ik}\|^2)}{\sum_{j=1}^K \exp(-s_j\|r_{ij}\|^2)}\,,
\end{equation}
where $s_k$ is a learnable smoothing factor for $c_k$.
Then the residual encoding for the \textit{k}-th codeword $c_k$ is calculated as follows:
\begin{equation}
e_{k} = \sum_{i=1}^N e_{ik} = \sum_{i=1}^N w_{ik}r_{ik}\,.
\end{equation}

In the speaker recognition task, Cai \textit{et al.} \cite{CaiLanguage} and Jung \textit{et al.} \cite{SPE} used the Encoding Layer for aggregating the frame-level speaker features to generate speaker embedding and improved the performance. 
Motivated by this, we apply the Encoding Layer in the speech quality assessment model. 
However, we use the Encoding Layer on frame-level scores, not frame-level feature vectors. 
Moreover, we utilize both GAP layer and Encoding Layer together to combine information from both layers.
In ground terrain recognition task, Xue \textit{et al.} \cite{DEP} showed it is better to use both the Encoding Layer and the GAP layer. 

Specifically, we put an Encoding Layer parallel to the GAP layer. 
The outputs from the Encoding Layer and the GAP layer are the residual encoding vector and the average score, respectively. 
They are concatenated and used as an input of the following FC layer to predict the utterance-level score.

When we use an Encoding Layer for similarity score prediction, the Encoding Layer aggregates the frame-level similarity scores as in MOS prediction. 

\section{Experiments}
\label{sec:experiments}

\subsection{Dataset}
As in \cite{MOSNet}, we use the MOS and similarity evaluation results from the Voice Conversion Challenge (VCC) 2018 \cite{VCC2018}. The challenge comprised two tasks: Hub task (parallel VC) and Spoke Task (non-parallel VC).
The VCC 2018 dataset is based on the device and production speech (DAPS) dataset \cite{Mysore2015}, which includes recordings of professional US English speakers. There are a total of eight source speakers and four target speakers. A total of 23 teams submitted systems to the Hub task, with 11 of them additionally participating in the Spoke task. There are a total of 38 evaluated systems, including the source speaker, the target speaker, and two baseline systems for the two tasks.

For MOS evaluation, 267 people rated the naturalness of 20,580 submitted utterances with a score ranging from 1 (``Completely unnatural") to 5 (``Completely natural"). The corresponding number of evaluation results is 82,304, and the ground-truth MOS of each utterance was obtained by averaging all the MOS ratings of the utterance. Among the 20,580 $<$audio, ground-truth MOS$>$ pairs, we use 15,580, 3,000, and 2,000 pairs for training, validation, and testing, respectively. 
The MOS of each system is obtained by averaging all the MOS values of the utterances from the system.

The same 267 people also rated the similarity between two utterances with a score among 1 (``Same, absolutely sure"), 2 (``Same, not sure"), 3 (``Different, not sure"), and 4 (``Different, absolutely sure"). 
An utterance pair consists of an anchor utterance, which can be either converted speech or human speech, and a reference utterance, which is an utterance from either the source or the target speaker of the anchor speech with the same linguistic context.
There are a total of 30,864 evaluation results, and the ground-truth similarity of each utterance pair is obtained from the average of the scores received. Among a total of 21,608 $<$audio pair, ground-truth similarity$>$ pairs, we use 17,286 for training, 2,161 for validation, and 2,161 for testing. We consider a pair $<$anchor system, reference system$>$ as one system pair, then the corresponding number of system pairs for similarity score prediction is 76.

We also use the MOS evaluation results of the VCC 2016 \cite{VCC2016} to test the generalization ability of the models trained on the VCC 2018 training set. The VCC 2016 comprised only a parallel voice conversion task, and there are a total of 26,028 utterances from 20 systems. Each system has 1,600 utterance-level evaluation results without any description of the utterances. 
Therefore, we can report only system-level performance on the evaluation results of the VCC 2016.

\subsection{Implementation details}
We implement all the models using PyTorch and train them on a single NVIDIA GTX 1080 Ti GPU with four different random seeds.
We set $\alpha$, the weighting factor for the frame-level MSE, to 0.8. 
When we use the GQT layer, we use 10 GQTs and 8 heads for a multi-head attention module. When we use the Encoding Layer, we use 10 codewords.
We use a batch size of 16 for MOSNet + GQT + EL and 32 for the rest of the models.
We use the Adam optimizer with a learning rate of 0.0001 and set a dropout rate to 0.3.
We use the validation set to select the model with the lowest MSE during 200 epochs.
We report the average MSE, linear correlation coefficient (LCC) \cite{Pearson}, and Spearman’s rank correlation coefficient (SRCC) \cite{Spearman} of the models trained with four random seeds. 

\begin{table}[t]
\caption{Results of different models. + GQT, + EL, + GQT + EL stand for using the GQT layer, the Encoding layer, and both of them, respectively. The best results are highlighted in bold.}
\label{tab:mos_results}
\vspace{-0.6cm}
\begin{scriptsize}
\begin{center}
\renewcommand{\tabcolsep}{0.76mm}
\renewcommand{\arraystretch}{1.03}
\begin{tabular}{l|cccccc|ccc}
\hline
\multicolumn{1}{c|}{\multirow{3}{*}{Model}}      & \multicolumn{6}{c|}{VCC 2018}                                            & \multicolumn{3}{c}{VCC 2016}  
\tabularnewline\cline{2-10}
\multicolumn{1}{c|}{}   & \multicolumn{3}{c}{\textit{utterance-level}} & \multicolumn{3}{c|}{\textit{system-level}} & \multicolumn{3}{c}{\textit{system-level}} \\
\multicolumn{1}{c|}{}    &    MSE        & LCC       & SRCC       & MSE       & LCC      & SRCC      & MSE       & LCC       & SRCC   \\\hline\hline
MOSNet      &   0.448   &  0.651   &  0.619    &  0.039   &  0.966  &  0.924   &  0.316   &  0.896   &  \textbf{0.858} \\
+ GQT        &    0.447  &  0.654   &  \textbf{0.621}    &  0.041   &  0.968  &   0.931  &  \textbf{0.242}   &  \textbf{0.921}   &  0.853\\
+ EL    &    \textbf{0.444}  &  \textbf{0.656}   &  0.617    &  \textbf{0.031}   &  \textbf{0.974}  &  0.938   &  \textbf{0.242}   &  0.908   &  0.855 \\
+ GQT + EL   &   0.447   &  0.656   &  0.616    &  0.032   &  0.967  &  \textbf{0.940}   &  0.246   &  0.885   &  0.839 \\
\hline
\end{tabular}
\end{center}
\end{scriptsize}
\vspace{-0.6cm}
\end{table}

\subsection{Experiments on MOS prediction}
First, we discuss the results on MOS prediction with the VCC 2018 test set, shown in Table \ref{tab:mos_results}.
MOSNet + GQT made improvements with all the metrics except the system-level MSE. To directly interpret the role of each GQT, we should adjust the weights of GQTs and observe the change in the predicted MOSs for various input utterances. However, this process is practically impossible since it requires listening to a lot of utterances and analyzing the factors that affect the MOS, which are expensive and subjective. 
Instead, from the MOS prediction results, we can infer that the GQTs become useful soft clusters for MOS evaluation.

Using the Encoding Layer improved all the metrics except the utterance-level SRCC as the model aggregated the frame-level scores using their distribution and learned the embeddings that are useful for the aggregation. It achieved the lowest MSE and highest LCC at both the utterance and system level. From the fact that MOSNet + EL shows better performance than MOSNet + GQT, we can say that considering the distribution of frame-level MOSs is more important than learning the quality embeddings for MOS evaluation.

When we combine MOSNet with both the GQT layer and Encoding Layer, the performance of MOSNet + GQT + EL is better than MOSNet + GQT but worse than MOSNet + EL. In other words, the Encoding Layer helps MOSNet + GQT, but the GQT layer does not help MOSNet + EL. 
As will be described in Section \ref{sec:embeddings}, we conjecture that GQTs prevent the embeddings of the utterances with similar scores from getting too far from each other, which results in that MOSNet + EL cannot separate the embeddings according to the frame-level scores as before.

With the test results using the VCC 2016 data, we can conclude that either the GQT layer or the Encoding Layer also improves the generalization ability of MOSNet. The Encoding Layer helps generalization through the sophisticated aggregation of frame-level scores. Moreover, the GQT layer directly improves the generalization ability of the model by learning the universal criteria for speech quality evaluation.

\subsection{Experiments on similarity score prediction}

The performance of the similarity score prediction models is evaluated with the MSE, LCC, SRCC, and accuracy, as shown in Table \ref{tab:similarity}. As we use a scalar for both the ground-truth and predicted similarity score, we regard the scores lower than 2.5 as the answer ``Same" and the scores higher or equal to 2.5 as the answer ``Different." Then we calculate the accuracy of a model as the ratio of cases when both answers from the model and humans are the same. 
All the proposed models show improvements in all the metrics except the accuracy.

Note that, unlike in MOS prediction, using the GQT layer shows better results than using the Encoding Layer.
We can infer that finding the criteria for voice similarity evaluation is more important than using the distribution of the frame-level similarity scores.
Furthermore, SIMNet + GQT + EL achieved the best performance among the four models, which means that SIMNet + EL takes advantage of the GQT layer. 
Considering this result and the fact that we judge the voice similarity throughout the whole utterance rather than specific frames, we can infer that SIMNet + EL does not separate the embeddings as far as in MOSNet + EL, according to the frame-level scores. 

Finally, we test how well our model, SIMNet + GQT + EL, approximates human assessment in terms of an evaluation method used in VCC 2018 \cite{VCC2018}. 
As mentioned earlier, the utterance-level similarity score can be classified into ``Same" or ``Different."
According to the method, the similarity score of a system is the ratio of the utterances that received ``Same" compared to the target speech.
Then we obtain the MSE, LCC, and SRCC by comparing the scores using the answers of our model and human, which are 0.037, 0.714, and 0.696, respectively.

\begin{table}[t]
\caption{Results of similarity prediction. ACC denotes accuracy. The best results are highlighted in bold.}
\label{tab:similarity}
\vspace{-0.62cm}
\begin{center}
\renewcommand{\tabcolsep}{1.2mm}
    \begin{tabular}{cccccc}
    Model & Level & MSE & LCC & SRCC & ACC\\
    \hline\hline
    \multirow{2}{*}{SIMNet} & utterance & 0.774 & 0.552 & 0.549 & 0.687 \tabularnewline
    & system &  0.052 & 0.925 & 0.905 & - \\ 
    \hline
    \multirow{2}{*}{+ GQT} & utterance & 0.763 & 0.558 & 0.554 & 0.687 \tabularnewline & system & 0.047 & 0.931 & 0.913 & - \\ 
    \hline
    \multirow{2}{*}{+ EL} & utterance & 0.770 & 0.555 & 0.554 & 0.684 \tabularnewline
    & system & 0.049 & 0.929 & 0.911 & - \\ 
    \hline
    \multirow{2}{*}{+ GQT + EL} & utterance & \textbf{0.761} & \textbf{0.560} & \textbf{0.558} & \textbf{0.689} \tabularnewline
    & system & \textbf{0.045} & \textbf{0.934} & \textbf{0.916} & - \\ 
    \hline
    \end{tabular}
\end{center}
\vspace{-0.5cm}
\end{table}

\begin{figure}[t] 
    \centering
    \includegraphics[trim=0.9cm 0.3cm 7.4cm 0.6cm, clip=true, width=8.0cm]{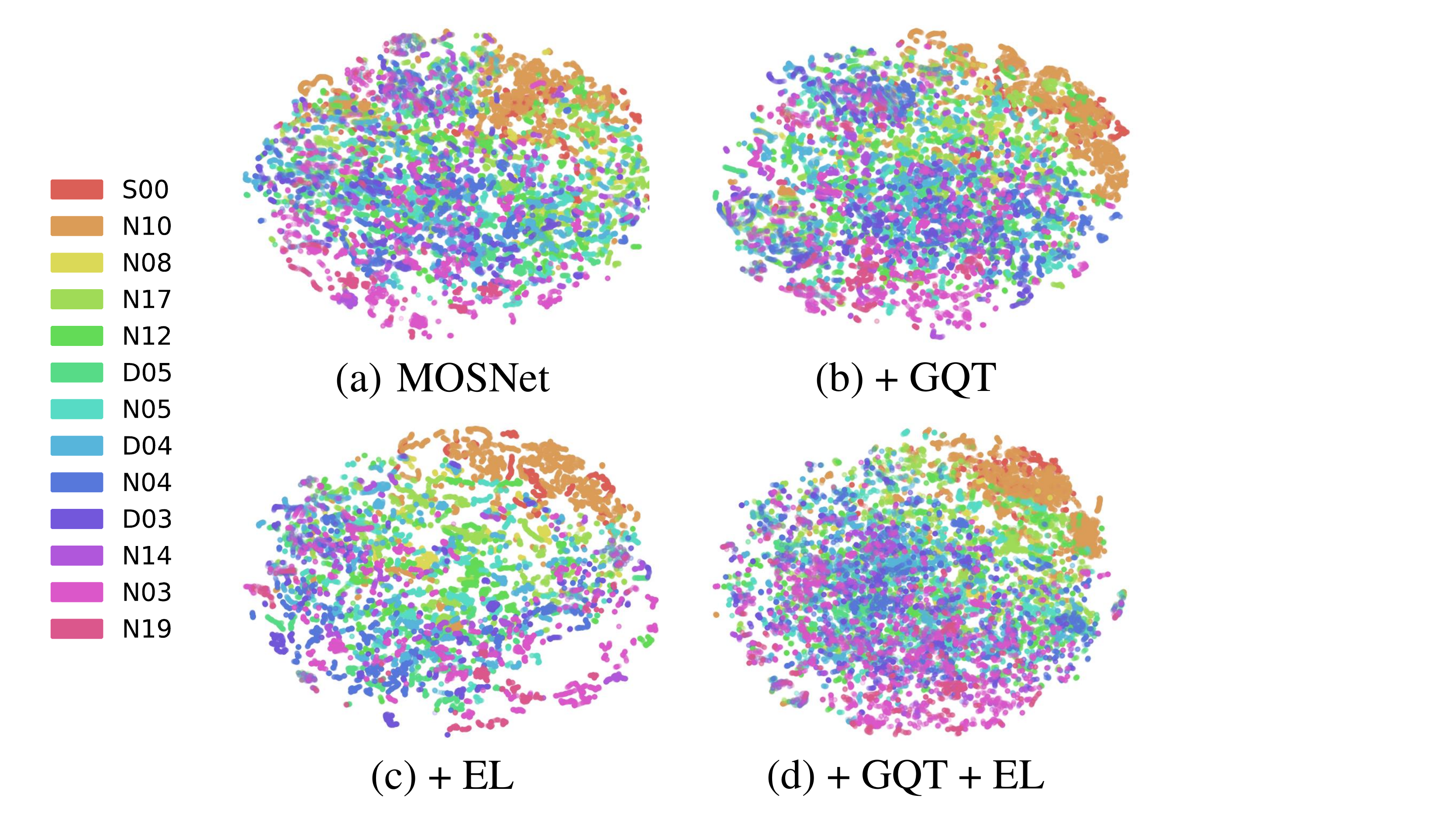}
\vspace{-0.3cm}
\caption{Visualization of embeddings for (a) MOSNet, (b) + GQT, (c) + EL, and (d) + GQT + EL}
\label{fig:tsne}
\vspace{-0.5cm}
\end{figure}

\subsection{Embeddings learned from the proposed models}
\label{sec:embeddings}
To discuss the effect of the GQT layer and Encoding Layer to the embeddings, we visualized the embeddings of four MOS prediction models using t-distributed stochastic neighbor embedding (t-SNE) \cite{tsne} in Figure \ref{fig:tsne}. 
We displayed a frame-level feature of CNN-BLSTM as a single dot.
We consider two systems submitted by one team as the same system because they are based on similar algorithms, which results in a total of 26 systems.
We use 390 random utterances from the VCC 2018 test set so that an average of 15 utterances per system exists.
We evenly select half of 26 systems after sorting the systems according to the MOS.
The color in Figure 1 indicates the system that generates the corresponding utterance.
Note that each system has its own system-level MOS in that different systems have different MOSs and the utterance-level MOSs are similar within each system.
The red dots are for the source speech, and the orange dots are for the voice conversion system with the highest MOS. 

Compared to (a), (b) shows that the GQT layer prevents the embeddings of the same system from getting apart from each other. 
Comparing (a) and (c), we see that the Encoding Layer separates the embeddings according to the system.
This shows that the Encoding Layer automatically learns the embeddings that are more useful for the aggregation, by utilizing the distribution of the frame-level scores.
In (d), we see that the embeddings of the same systems get close to each other while embeddings of different systems get apart from each other. 
Specifically, + GQT + EL learns distinguishable embeddings between different systems that have higher MOSs, including S00, N10, and N17. 
However, the other systems having lower MOSs are less distinguishable from each other than in MOS + EL. We infer that this is the main reason for the degradation of performance.

\section{Conclusion}
\label{sec:conclusion}
We proposed three deep-learning-based speech quality assessment models using cluster-based modeling, which improved MOSNet and SIMNet using a GQT layer, an Encoding Layer, and both of them, respectively. With experimental results on MOS and similarity score prediction, we showed that the GQT layer learns the criteria of speech quality evaluation as soft clusters, and the Encoding Layer utilizes the frame-level scores in a more sophisticated way.
For future work, we will apply our models to approximate other speech quality assessments, such as PESQ.
Furthermore, we will use our models to guide current TTS models to learn human perception, by using a perceptual loss in training.
Finally, we will figure out how to create the synergy between using the GQT layer and using the Encoding Layer for MOS prediction.

\section{Acknowledgements}

This material is based upon work supported by the Ministry of Trade, Industry \& Energy (MOTIE, Korea) under Industrial Technology Innovation Program (No. 10080667, Development of conversational speech synthesis technology to express emotion and personality of robots through sound source diversification).

\bibliographystyle{IEEEtran}

\bibliography{template}

\begin{thebibliography}{10}
\providecommand{\url}[1]{#1}
\csname url@samestyle\endcsname
\providecommand{\newblock}{\relax}
\providecommand{\bibinfo}[2]{#2}
\providecommand{\BIBentrySTDinterwordspacing}{\spaceskip=0pt\relax}
\providecommand{\BIBentryALTinterwordstretchfactor}{4}
\providecommand{\BIBentryALTinterwordspacing}{\spaceskip=\fontdimen2\font plus
\BIBentryALTinterwordstretchfactor\fontdimen3\font minus
  \fontdimen4\font\relax}
\providecommand{\BIBforeignlanguage}[2]{{%
\expandafter\ifx\csname l@#1\endcsname\relax
\typeout{** WARNING: IEEEtran.bst: No hyphenation pattern has been}%
\typeout{** loaded for the language `#1'. Using the pattern for}%
\typeout{** the default language instead.}%
\else
\language=\csname l@#1\endcsname
\fi
#2}}
\providecommand{\BIBdecl}{\relax}
\BIBdecl

\bibitem{Han2019}
K.~J. Han, R.~Prieto, K.~Wu, and T.~Ma, ``State-of-the-art speech recognition
  using multi-stream self-attention with dilated 1d convolutions,'' \emph{arXiv
  preprint arXiv:1910.00716}, 2019.

\bibitem{WaveNet}
A.~Oord, S.~Dieleman, H.~Zen, K.~Simonyan, O.~Vinyals, A.~Graves,
  N.~Kalchbrenner, A.~Senior, and K.~Kavukcuoglu, ``Wave{N}et: A generative
  model for raw audio,'' \emph{arXiv preprint arXiv:1609.03499}, 2016.

\bibitem{Tacotron2}
J.~Shen, R.~Pang, R.~J. Weiss, M.~Schuster, N.~Jaitly, Z.~Yang, Z.~Chen,
  Y.~Zhang, Y.~Wang, R.~Skerry-Ryan, R.~A. Saurous, Y.~Agiomyrgiannakis, and
  Y.~Wu, ``Natural {TTS} synthesis by conditioning {WaveNet} on {M}el
  spectrogram predictions,'' in \emph{Proc. of the IEEE International
  Conference on Acoustics, Speech and Signal Processing (ICASSP)}, 2018, pp.
  4779--4783.

\bibitem{TransformerTTS}
N.~Li, S.~Liu, Y.~Liu, S.~Zhao, and M.~Liu, ``Neural speech synthesis with
  {T}ransformer network,'' in \emph{Proc. of the AAAI Conference on Artificial
  Intelligence}, 2019, pp. 6706--6713.

\bibitem{King}
S.~King, ``Measuring a decade of progress in text-to-speech,'' \emph{Loquens},
  vol.~1, no.~1, 2014.

\bibitem{Eval}
P.~Wagner, J.~Beskow, S.~Betz, J.~Edlund, J.~Gustafson, G.~E. Henter, S.~L.
  Maguer, Z.~Malisz, E.~Sz{\'{e}}kely, C.~T{\aa}nnander, and J.~Vo{\ss}e,
  ``Speech synthesis evaluation — state-of-the-art assessment and suggestion
  for a novel research program,'' in \emph{Proc. 10th Speech Synthesis Workshop
  (SSW10)}, 2019.

\bibitem{MCD}
R.~Kubichek, ``Mel-cepstral distance measure for objective speech quality
  assessment,'' in \emph{Proc. IEEE Pacific Rim Conference on Communications
  Computers and Signal Processing}, 1993, pp. 125--128.

\bibitem{PESQ}
A.~W. Rix, J.~G. Beerends, M.~P. Hollier, and A.~P. Hekstra, ``Perceptual
  evaluation of speech quality ({PESQ})-a new method for speech quality
  assessment of telephone networks and codecs,'' in \emph{Proc. of the IEEE
  International Conference on Acoustics, Speech and Signal Processing
  (ICASSP)}, 2001, pp. 749--752.

\bibitem{ANIQUE}
D.-S. Kim, ``{ANIQUE}: an auditory model for single-ended speech quality
  estimation,'' \emph{IEEE Transactions on Speech and Audio Processing},
  vol.~13, pp. 821--831, 2005.

\bibitem{P563}
L.~Malfait, J.~Berger, and M.~Kastner, ``P. 563: The {ITU-T} standard for
  single-ended speech quality assessment,'' \emph{IEEE Transactions on Audio,
  Speech, and Language Processing}, vol.~14, pp. 1924--1934, 2006.

\bibitem{HierarchicalAssessment}
T.~Yoshimura, G.~E. Henter, O.~Watts, M.~Wester, J.~Yamagishi, and K.~Tokuda,
  ``A hierarchical predictor of synthetic speech naturalness using neural
  networks,'' in \emph{Proc. of Interspeech}, 2016, pp. 342--346.

\bibitem{AutoMOS}
B.~Patton, Y.~Agiomyrgiannakis, M.~Terry, K.~W. Wilson, R.~A. Saurous, and
  D.~Sculley, ``Auto{MOS}: Learning a non-intrusive assessor of
  naturalness-of-speech,'' in \emph{Proc. of NIPS End-to-end Learning for
  Speech and Audio Processing Workshop}, 2016.

\bibitem{QualityNet}
S.~Fu, Y.~Tsao, H.~Hwang, and H.~Wang, ``Quality-{N}et: An end-to-end
  non-intrusive speech quality assessment model based on blstm,'' in
  \emph{Proc. of Interspeech}, 2018, pp. 1873--1877.

\bibitem{MOSNet}
C.~Lo, S.~Fu, W.~Huang, X.~Wang, J.~Yamagishi, Y.~Tsao, and H.~Wang,
  ``{MOSNet}: Deep learning based objective assessment for voice conversion,''
  in \emph{Proc. of Interspeech}, 2019, pp. 1541--1545.

\bibitem{GST}
Y.~Wang, D.~Stanton, Y.~Zhang, R.~Skerry-Ryan, E.~Battenberg, J.~Shor, Y.~Xiao,
  F.~Ren, Y.~Jia, and R.~A. Saurous, ``Style tokens: Unsupervised style
  modeling, control and transfer in end-to-end speech synthesis,'' \emph{arXiv
  preprint arXiv:1803.09017}, 2018.

\bibitem{AttentionIsAllYouNeed}
A.~Vaswani, N.~Shazeer, N.~Parmar, J.~Uszkoreit, L.~Jones, A.~N. Gomez,
  L.~Kaiser, and I.~Polosukhin, ``Attention is all you need,'' in
  \emph{Advances in Neural Information Processing Systems (NIPS)}, 2017, pp.
  6000--6010.

\bibitem{ImprovingUnsupervisedStyleTransfer}
D.~Liu, C.~Yang, S.~Wu, and H.~Lee, ``Improving unsupervised style transfer in
  end-to-end speech synthesis with end-to-end speech recognition,'' in
  \emph{Proc. of Spoken Language Technology Workshop (SLT)}, 2018, pp.
  640--647.

\bibitem{GRU}
K.~Cho, B.~Merrienboer, C.~Gulcehre, D.~Bahdanau, F.~Bougares, H.~Schwenk, and
  Y.~Bengio, ``Learning phrase representations using {RNN} encoder-decoder for
  statistical machine translation,'' in \emph{Proc. of Empirical Methods in
  Natural Language Processing (EMNLP)}, 2014, pp. 1724--1734.

\bibitem{DeepTEN}
H.~Zhang, J.~Xue, and K.~Dana, ``Deep ten: Texture encoding network,'' in
  \emph{Proc. of Computer Vision and Pattern Recognition (CVPR)}, 2017, pp.
  708--717.

\bibitem{CaiLanguage}
W.~Cai, Z.~Cai, X.~Zhang, X.~Wang, and M.~Li, ``A novel learnable dictionary
  encoding layer for end-to-end language identification,'' in \emph{Proc. of
  the IEEE International Conference on Acoustics, Speech and Signal Processing
  (ICASSP)}, 2018, pp. 5189--5193.

\bibitem{SPE}
Y.~Jung, Y.~Kim, H.~Lim, Y.~Choi, and H.~Kim, ``Spatial pyramid encoding with
  convex length normalization for text-independent speaker verification,'' in
  \emph{Proc. of Interspeech}, 2019, pp. 4030--4034.

\bibitem{DEP}
J.~Xue, H.~Zhang, and K.~Dana, ``Deep texture manifold for ground terrain
  recognition,'' in \emph{Proc. of Computer Vision and Pattern Recognition
  (CVPR)}, 2018, pp. 558--567.

\bibitem{VCC2018}
J.~Lorenzo-Trueba, J.~Yamagishi, T.~Toda, D.~Saito, F.~Villavicencio,
  T.~Kinnunen, and Z.~Ling, ``The voice conversion challenge 2018: Promoting
  development of parallel and nonparallel methods,'' in \emph{Proc. of Odyssey
  The Speaker and Language Recognition Workshop}, 2018, pp. 195--202.

\bibitem{Mysore2015}
G.~J. Mysore, ``Can we automatically transform speech recorded on common
  consumer devices in real-world environments into professional production
  quality speech?—a dataset, insights, and challenges,'' \emph{IEEE Signal
  Processing Letters}, vol.~22, no.~8, pp. 1006--1010, 2015.

\bibitem{VCC2016}
T.~Toda, L.~Chen, D.~Saito, F.~Villavicencio, M.~Wester, Z.~Wu, and
  J.~Yamagishi, ``The {V}oice {C}onversion {C}hallenge 2016,'' in \emph{Proc.
  of Interspeech}, 2016, pp. 1632--1636.

\bibitem{Pearson}
K.~Pearson, ``Notes on the history of correlation,'' \emph{Biometrika},
  vol.~13, no.~1, pp. 25--45, 1920.

\bibitem{Spearman}
C.~Spearman, ``The proof and measurement of association between two things,''
  \emph{The American Journal of Psychology}, vol.~15, no.~1, pp. 72--101, 1904.

\bibitem{tsne}
L.~V.~D. Maaten and G.~Hinton, ``Visualizing data using t-{SNE},''
  \emph{Journal of machine learning research}, no.~9, pp. 2579--2605, 2008.

\end{thebibliography}

\end{document}